\documentclass[twocolumn,fleqn]{article}
\usepackage[utf8]{inputenc}
\usepackage[T1]{fontenc}
\usepackage{amsmath,amssymb}
\usepackage{geometry}
\usepackage{hyperref}
\title{SEMI-EMPIRICAL KINETIC MODEL FOR PHASE SELECTION IN RAPIDLY SOLIDIFIED MULTICOMPONENT CONCENTRATED ALLOYS}

\author{%
O.I.~Kushnerov\\
Oles Honchar Dnipro National University, 72, Gagarin Ave., Dnipro 49045, Ukraine\\
\texttt{kushnrv@gmail.com}
\and
S.I.~Ryabtsev\\
Oles Honchar Dnipro National University, 72, Gagarin Ave., Dnipro 49045, Ukraine
\and
V.F.~Bashev\\
Dniprovsky State Technical University, 2, Dniprobudivska Str., Kamianske 51918, Ukraine
}
\date{Preprint version}

\begin{document}
\twocolumn[
\maketitle
\begin{abstract}
A semi-empirical kinetic framework is formulated for predicting phase selection in multicomponent concentrated alloys under rapid solidification. The approach is based on the critical cooling rate required to suppress competing crystalline pathways and combines topology-dependent ranking of BCC-, FCC-, and HCP-like crystallization pathways with a correction for glass-forming ability. The formulation includes a topology-dependent viscosity correction with a smoothed BCC multiplier and a continuous correction factor for glass-forming ability based on mixing enthalpy, excess entropy, and atomic-size dispersion. Comparison with experimental and computational data shows that the kinetic criterion captures changes in the lattice type expected from the valence electron concentration criterion, describes kinetic suppression of phase separation, and identifies competitive multiphase crystallization. The model also distinguishes alloys with high and low glass-forming ability. The proposed framework provides a practical approach for preliminary evaluation of kinetic phase competition in rapidly solidified multicomponent melts.
\end{abstract}
\vspace{0.5em}
\vspace{1.0em}
]

\section{Introduction}

Multicomponent concentrated alloys form a broad family of metallic systems described by closely related and partly overlapping terms, including high-entropy alloys (HEAs), complex concentrated alloys (CCAs), and multi-principal-element alloys (MPEAs). These terms emphasize different aspects of the same compositional concept, namely high configurational entropy, compositional complexity, or the presence of several principal elements. In such systems, phase formation and properties are governed by chemical, topological, electronic, and kinetic factors. Stimulated by the pioneering works of Yeh et al.~\cite{Yeh2004} and Cantor et al.~\cite{Cantor2004}, this compositional design strategy has transformed modern physical metallurgy by shifting alloy design from single-principal-element systems to compositions containing several principal elements and by opening new possibilities for controlling phase formation, structure--property relationships, functional behavior, and applications~\cite{Biswas2022,Xiang2023,Zhang2025FunctionalHEA}.

The growing interest in these materials is associated with their ability to combine high strength, hardness, thermal stability, wear and corrosion resistance, and, in selected systems, functional properties such as superconductivity, shape-memory behavior, or enhanced radiation tolerance~\cite{Miracle2017,Wang2014,KitagawaNbScTiZr2024,Koval2025ShapeMemory,NahavandianIrradiation2025}. At the same time, the enormous compositional space of multicomponent alloys has stimulated the development of data-driven and machine-learning approaches for accelerated materials discovery~\cite{ZhangWordEmbeddings2025}.

Multicomponent concentrated alloys can be produced by various processing routes, including arc melting, powder metallurgy, mechanical alloying, additive manufacturing, thin-film deposition, and rapid solidification techniques. Among these methods, rapid quenching, laser alloying, and thin-film deposition are especially important because they impose strongly non-equilibrium solidification or growth conditions. In rapid quenching and laser alloying, cooling rates may exceed $10^4$~K/s, whereas thin-film deposition additionally restricts atomic diffusion through nanoscale growth geometry and high heat extraction rates. Under such conditions, diffusional partitioning can be suppressed, the microstructure is refined, and metastable solid solutions, nanocrystalline structures, or amorphous states may form~\cite{KushnerovStaticCriteria,Kushnerov2021CoCrCuFeNi,Kushnerov2023MetastableFilms,Dekhtyarenko2025Amorphous,Girzhon2023AMS,Girzhon2025UFM}. Such effects were observed, for example, in splat-quenched CoCr$_{0.8}$Cu$_{0.64}$FeNi and in metastable Co--Cr--Fe--Mn--Ni high-entropy alloy thin films, where rapid cooling or non-equilibrium deposition changed the structural state and physical properties of the alloys~\cite{Kushnerov2021CoCrCuFeNi,Kushnerov2023MetastableFilms}. Therefore, these non-equilibrium processing routes represent not only technological methods for obtaining improved properties, but also physical regimes in which phase selection must be treated as a kinetic rather than purely equilibrium problem.

Numerous empirical and semi-empirical criteria have been proposed to rationalize phase formation in these systems. Most of them are based on static parameters, including the valence electron concentration, the enthalpy and entropy of mixing, the atomic size mismatch, electronegativity differences, and related thermodynamic or topological parameters~\cite{Miracle2017,Yang2012,Wang2014}. These criteria are useful under near-equilibrium conditions, but their predictive capability becomes limited when alloys are produced by rapid solidification.

Under rapid cooling, where effective cooling rates may reach or exceed $10^6$~K/s, phase selection is governed not only by thermodynamic stability, but also by kinetic competition between alternative nucleation and growth pathways. The first phase to appear may therefore be not the equilibrium phase, but the structure with the lowest effective kinetic barrier for nucleation. A kinetic description is consequently required to account for atomic mobility, melt viscosity, incubation time, and the finite time available for crystallization.

Classical nucleation models provide the physical basis for such an approach, but their direct application to multicomponent concentrated alloys is limited by the lack of reliable data on interfacial energies, wetting angles, diffusion coefficients, and nucleant characteristics. A more practical route is based on the critical cooling rate required to suppress crystallization. The model of Takeuchi and Inoue relates this quantity to the thermodynamic stability of the liquid phase and the effective viscosity of the melt~\cite{Takeuchi2001,Takeuchi2005}. However, its application to complex concentrated alloys requires a consistent treatment of topology-dependent viscosity and glass-forming ability.

In previous studies, the thermodynamic, electronic, atomic-size, and topological parameters of the investigated alloy systems were systematically evaluated and compared with experimentally observed phase constitutions~\cite{KushnerovStaticCriteria}. These experimental and parameter datasets provide the quantitative basis for the present work, in which the critical cooling rates for competing BCC-, FCC-, and HCP-like crystallization pathways are evaluated within a regularized semi-empirical kinetic framework. The present study focuses on the physical basis, mathematical regularization, glass-forming ability correction, and applicability limits of the proposed phase-selection criterion.

The objective of this work is to formulate a compact semi-empirical kinetic framework for predicting phase formation in rapidly solidified multicomponent concentrated alloys, with particular emphasis on the regularization of the BCC topological multiplier, the incorporation of a glass-forming ability correction, and the applicability limits of the resulting phase-selection criterion.

The proposed approach is not intended to replace CALPHAD calculations, atomistic simulations, or experimental phase analysis. It is designed for preliminary evaluation of kinetic phase competition in chemically complex melts, for which complete information on interfacial energies, diffusion coefficients, wetting conditions, and nucleation-site characteristics is usually unavailable. In this sense, the model provides a practical framework for identifying probable kinetically dominant crystallization pathways under rapid solidification conditions.

\section{Model Formulation}

The prediction of phase formation in multicomponent concentrated alloys under non-equilibrium solidification requires a framework that consistently couples thermodynamic driving forces with kinetic limitations. Conventional empirical criteria provide useful estimates of phase stability under near-equilibrium conditions, but they inherently neglect atomic mobility and the time-dependent nature of phase transformations. To overcome these limitations, a predictive model for rapidly solidified multicomponent alloys must combine thermodynamic parameters with kinetic parameters that remain calculable for chemically complex systems.

\subsection{Critical Cooling Rate and Topology-Dependent Viscosity}

A fundamental description of crystallization in undercooled melts is provided by classical nucleation theory. In kinetic formalisms developed for homogeneous and heterogeneous nucleation, the characteristic incubation time is controlled by melt viscosity, diffusion mobility, thermodynamic driving force, interfacial conditions, and the degree of undercooling~\cite{Uhlmann1972,Davies1976,ShaoTsakiropoulos1994}. For high-entropy alloys, a kinetic approach to phase prediction was developed by Chattopadhyay et al.~\cite{Chattopadhyay2018}, who considered competition between FCC-, BCC-, and HCP-like crystallization pathways using melt viscosity and structural information. At the same time, the non-stationary approach of Girzhon, Smolyakov, and Yemelianchenko~\cite{GirzhonSmolyakov,Girzhon2023AMS,Girzhon2025UFM} explicitly treats heterogeneous nucleation under rapid-cooling conditions.

In real rapid-solidification experiments, several nucleation mechanisms may coexist, including heterogeneous nucleation at the melt--substrate interface, nucleation on internal catalytic sites, and homogeneous nucleation in the melt volume at sufficiently large undercooling. Therefore, in the present work, as in the Takeuchi--Inoue model~\cite{Takeuchi2001}, no specific microscopic nucleation mechanism is prescribed. Instead, the kinetic barrier is described using an effective macroscopic parameter, namely the critical cooling rate required to suppress the corresponding crystallization pathway. In the model of Takeuchi and Inoue, the critical cooling rate is related to the thermodynamic stability of the liquid phase and the effective viscosity of the melt~\cite{Takeuchi2001}:

\begin{equation}
	R_c =
	Z
	\frac{k_B T_m^2}{a^3 \eta(T_m)}
	\exp\left(
	\frac{\Delta G_{mix}}{R T}
	\right),
\end{equation}

where $Z$ is a dimensionless empirical constant, $k_B$ is the Boltzmann constant, $T_m$ is the melting temperature, $a$ is the average interatomic distance, $\eta(T_m)$ is the dynamic viscosity at $T_m$, $R$ is the universal gas constant, $\Delta G_{mix}$ is the Gibbs free energy of mixing, and $T$ is the temperature used in the thermodynamic correction to the kinetic prefactor. The pre-exponential factor describes the kinetic contribution through viscosity-dependent atomic mobility near the melting temperature, whereas the exponential term accounts for the thermodynamic stabilization of the liquid state due to mixing.

For practical application to multicomponent alloys, the Gibbs free energy of mixing is expressed as

\begin{equation}
	\Delta G_{mix}
	=
	\Delta H_{mix}
	-
	T_m \Delta S_{res},
\end{equation}

where $\Delta H_{mix}$ is the enthalpy of mixing and $\Delta S_{res}$ is the residual entropy contribution. In the present framework, this entropy term is treated as

\begin{equation}
	\Delta S_{res}
	=
	\Delta S_{mix}
	+
	\Delta S_{xs},
\end{equation}

where $\Delta S_{mix}$ is the ideal configurational entropy of mixing and $\Delta S_{xs}$ is the excess entropy associated with non-ideal packing and atomic-size mismatch. Other entropy contributions are generally unavailable for most multicomponent melts and are not separated explicitly in the present semi-empirical treatment. Their averaged impact is partially taken into account by calibration coefficients.

The central quantity controlling the kinetic pre-exponential factor is the melt viscosity $\eta(T_m)$. Experimental studies of Cu-based binary subsystems of the Bi--Cu--Ga--Sn--Pb high-entropy alloy have shown that viscosity is a structure-sensitive property of multicomponent metallic melts and that negative mixing entropy values can indicate structural ordering in the liquid state~\cite{Dufanets2020}. In chemically complex alloys, the effective viscosity is therefore not only a function of temperature and composition but may also be sensitive to the local atomic topology that precedes crystallization. Following the kinetic interpretation proposed by Chattopadhyay et al.~\cite{Chattopadhyay2018}, the undercooled melt is assumed to contain short-range structural fluctuations associated with competing FCC-, BCC-, and HCP-like crystallization pathways. To estimate the corresponding effective viscosity, a modified relation based on the approach of Budai et al.~\cite{Budai2007} was used:

\begin{equation}
	\eta(T)=
	P(\delta r)
	\frac{hN_A}{V_{alloy}}
	\exp[\Psi_\eta(T)],
\end{equation}

\begin{equation}
	\begin{split}
		\Psi_\eta(T) ={}&
		\sum_{i=1}^{n}c_i\ln\eta_i
		+
		\ln\left(\frac{V_c}{hN_c}\right)
		\\
		&{}-
		\frac{0.155\,\Delta H_{mix}}{RT}.
	\end{split}
\end{equation}

Here $h$ is the Planck constant, $N_A$ is the Avogadro number, $V_{alloy}$ is the molar volume of the alloy, $c_i$ is the atomic fraction of the $i$-th element, $\eta_i$ is the viscosity of the individual element, $V_c$ is the volume of the candidate crystalline cell, $N_c$ is the effective number of atoms in this cell, and $P(\delta r)$ is the topology-dependent multiplier.

The molar volume of the alloy was calculated from the molar volumes of the constituent elements using the rule of mixtures,

\begin{equation}
	V_{alloy}=\sum_{i=1}^{n}c_i V_i ,
\end{equation}

where $V_i$ is the molar volume of the $i$-th element. The volume of the candidate cell, $V_c$, was determined using the hard-sphere geometric approximation,

\begin{equation}
	V_c=p\bar r^3,
\end{equation}

where $\bar r$ is the composition-averaged atomic radius and $p$ is a structure-dependent geometric constant. To justify the numerical values of $p$, the relationship between the cell volume, the number of atoms in the cell, and the atomic packing factor $\epsilon$ can be written as

\begin{equation}
	V_c=
	\frac{
		N_c(4\pi/3)\bar r^3
	}{
		\epsilon
	}.
\end{equation}

Equating the two expressions gives $p=4\pi N_c/(3\epsilon)$. For the BCC lattice, $N_c=2$ and $\epsilon\simeq0.68$, giving $p\simeq12.32$. For the FCC lattice, $N_c=4$ and $\epsilon\simeq0.74$, giving $p\simeq22.63$. For the HCP lattice, the present formulation uses the full conventional hexagonal cell, i.e. a full hexagonal prism containing six atoms. In this convention, $N_c=6$ and $\epsilon\simeq0.74$, giving $p\simeq33.94$. 
This convention is equivalent to using the primitive HCP cell with $N_c=2$ and $p\simeq11.31$, because the viscosity expression contains the ratio $V_c/N_c$.

To ensure the applicability of the kinetic model to non-equiatomic concentrated alloys, the atomic-size variation parameter was written in the mole-fraction-weighted form. This quantity is mathematically equivalent to the classical atomic-size mismatch parameter $\delta r$ widely used in high-entropy alloy physics:

\begin{equation}
	\delta r =
	\left[
	\sum_{i=1}^{n}
	c_i
	\left(
	1-\frac{r_i}{\bar r}
	\right)^2
	\right]^{1/2},
	\qquad
	\bar r =
	\sum_{i=1}^{n} c_i r_i ,
\end{equation}

where $r_i$ is the atomic radius of the $i$-th element and $c_i$ is its atomic fraction. This formulation accounts for the proportional contribution of both principal elements and minor alloying additions to the overall topological distortion of atomic packing and connects the kinetic model with conventional static parameters of atomic-size mismatch.

The individual elemental viscosities $\eta_i(T)$ were evaluated using the Arrhenius approximation. Although the Vogel--Fulcher--Tammann relation is more appropriate for describing the strongly non-linear viscosity increase in deeply undercooled glass-forming liquids, the Arrhenius form provides a practical estimate of the baseline elemental mobility near the melting temperature. In the present semi-empirical framework, additional kinetic slowdown associated with atomic-size mismatch and glass-forming frustration is introduced through the topology-dependent multiplier $P(\delta r)$ and the glass-forming ability correction factor.

The multiplier $P(\delta r)$ accounts for differences in atomic packing efficiency and local structural order for competing crystal lattices. For closely packed structures, such as face-centered cubic and hexagonal close-packed lattices, the topological multiplier is defined as a linear function,

\begin{equation}
	P_{FCC/HCP}(\delta r)=1+\epsilon N_c \delta r .
\end{equation}

Here $\epsilon$ is the atomic packing factor of the corresponding lattice. In the present formulation, $\epsilon=0.74$ for FCC and HCP structures, while $N_c=4$ for FCC and $N_c=6$ for HCP. The use of $N_c=6$ for HCP corresponds to the full conventional hexagonal prism considered in the topological multiplier.

For the less densely packed BCC structure, the formulation used in earlier kinetic approaches is piecewise-defined:

\begin{equation}
	P_{BCC}=1+\frac{\epsilon N_c}{\delta r},
	\quad
	0<\delta r<0.06,
\end{equation}

\begin{equation}
	P_{BCC}=1+\epsilon N_c \delta r,
	\quad
	\delta r\geq0.06,
\end{equation}

where $\epsilon=0.68$ and $N_c=2$ for the BCC lattice. This piecewise form reflects the assumption that the kinetic preference for BCC-like ordering changes when atomic-size dispersion exceeds a critical threshold. However, the abrupt transition between two analytical branches produces a non-physical jump in the calculated topological multiplier and, consequently, in the effective viscosity.

Since the critical cooling rate depends sensitively on viscosity, this discontinuity can generate artificial jumps in the calculated $R_c$ values and may distort the predicted phase ranking when BCC, FCC, and HCP pathways have close kinetic priorities. To remove this artifact, the discontinuous switch between the two BCC branches is replaced by a continuous weighted interpolation.

The transition region was shifted to $\delta r\approx0.05$ to provide smooth matching of the low- and high-mismatch regions before reaching the critical region near $\delta r\approx0.06$. The weighting functions are defined as

\begin{equation}
	W_l(\delta r)=
	\frac{1}{2}
	\left[
	1-\tanh\left(500(\delta r-0.05)\right)
	\right],
\end{equation}

\begin{equation}
	W_h(\delta r)=
	\frac{1}{2}
	\left[
	1+\tanh\left(500(\delta r-0.05)\right)
	\right].
\end{equation}

The function $W_l(\delta r)$ gives the weight of the low atomic-size-mismatch region, whereas $W_h(\delta r)$ gives the weight of the high atomic-size-mismatch region. These functions satisfy

\begin{equation}
	W_l(\delta r)+W_h(\delta r)=1,
\end{equation}

which ensures a continuous weighted matching of the two limiting analytical branches. The resulting continuous BCC topological multiplier is written as

\begin{equation}
	\begin{split}
		P_{BCC}(\delta r)={}&
		W_l(\delta r)
		\left[
		18.61
		\left(
		\delta r+
		\frac{6.25\times10^{-5}}{\delta r^2}
		\right)
		\right]
		\\
		&{}+
		W_h(\delta r)
		(0.895+0.1\delta r).
	\end{split}
\end{equation}

The coefficients in the regularized BCC multiplier were selected by numerical parametric testing to ensure continuity, boundedness, and stable viscosity calculations in the transition region. They are not treated as new universal physical constants, but as semi-empirical regularization parameters that preserve the limiting behavior of the two original BCC branches while removing the non-physical discontinuity.
This expression should be regarded as a regularized interpolation between the two limiting BCC branches. The hyperbolic-tangent weighting functions provide a compact monotonic transition with controllable width and preserve the asymptotic behavior of the original branches outside the transition region. As a result, the calculation of the effective viscosity is stabilized, and differences in the calculated critical cooling rates originate from the thermodynamic and topological characteristics of the alloy rather than from a discontinuity in the mathematical representation.

\subsection{Glass-Forming Ability Correction and Phase-Selection Criterion}

The topology-dependent viscosity correction stabilizes the relative assessment of competing crystalline pathways. However, the original macroscopic critical-cooling-rate model may still yield overly optimistic, i.e. underestimated, values of $R_c$ for alloys that are intrinsically poor glass formers. This limitation arises because the thermodynamic exponential term does not fully capture the threshold character of empirical glass-forming rules. For example, an alloy with a slightly negative enthalpy of mixing may formally exhibit a reduced calculated $R_c$, although such a composition may still be difficult to amorphize experimentally.

To account for this effect, a glass-forming ability correction factor $F_{GFA}$ is introduced. The correction is based on a continuous glass-forming ability index, $GFA_{index}$, which combines three smooth threshold scores describing the influence of mixing enthalpy, excess entropy, and atomic-size mismatch:

\begin{equation}
	\begin{aligned}
		Score_H &=
		1-
		\frac{1}{
			1+\exp[-k_H(\Delta H_{mix}-H_0)]
		},
		\\
		Score_S &=
		\frac{1}{
			1+\exp[-k_S(\Delta S_{xs}/R-S_0)]
		},
		\\
		Score_d &=
		\frac{1}{
			1+\exp[-k_d(\delta r-\delta_0)]
		}.
	\end{aligned}
\end{equation}

Here $Score_H$ reflects the requirement of sufficiently negative mixing enthalpy, $Score_S$ describes the contribution of non-ideal packing through the normalized excess entropy, and $Score_d$ accounts directly for atomic-size mismatch. The characteristic threshold values corresponding to the midpoints of the smooth threshold transitions are $H_0=-15$~kJ/mol, $S_0=0.1$, and $\delta_0=0.08$, while $k_H=1$ and $k_S=k_d=100$ determine the sharpness of these transitions. These threshold values are consistent with empirical parameters commonly used to distinguish solid-solution, intermetallic, and amorphous-forming tendencies in multicomponent metallic systems~\cite{Zhang2008,Yang2012,Mansoori1971,Takeuchi2005}.

The three scores are combined multiplicatively,

\begin{equation}
	GFA_{index}
	=
	Score_H Score_S Score_d.
\end{equation}

The multiplicative form ensures that a high $GFA_{index}$ is obtained only when the enthalpic, entropic, and atomic-size conditions favorable for glass formation are satisfied simultaneously. Thus, $GFA_{index}$ approaches unity only for alloys in which all three contributions support amorphization. The final correction factor is defined as

\begin{equation}
	F_{GFA}
	=
	P_{max}^{(1-GFA_{index})},
\end{equation}

where $P_{max}=30$ defines the upper limit of the correction multiplier. This value was selected to keep the limiting calculated critical cooling rates in the range characteristic of metallic systems with low glass-forming ability, including pure transition metals such as Ni. When $GFA_{index}\rightarrow1$, the correction approaches unity, corresponding to high glass-forming ability. When $GFA_{index}\rightarrow0$, the calculated critical cooling rate is increased, reflecting the higher cooling rate required to suppress crystallization in alloys with low glass-forming ability.

By combining the topology-dependent viscosity correction and the glass-forming ability correction factor, the final semi-empirical expression for the critical cooling rate is written as

\begin{equation}
	R_c =
	Z
	\frac{k_B T_m^2}{a^3\eta(T_m)}
	\exp(\Phi)
	F_{GFA},
\end{equation}

\begin{equation}
	\begin{split}
		\Phi ={}&
		f_1
		\frac{\Delta H_{mix}-T_m\Delta S_{mix}}{R T_{ref}}
		\\
		&{}-
		f_2
		\frac{T_m\Delta S_{xs}}{R T_{ref}} .
	\end{split}
\end{equation}

Here $T_{ref}=300$~K is the reference temperature. It corresponds approximately to the final room-temperature state of the cooled samples and is used to normalize the thermodynamic correction terms. The kinetic prefactor remains controlled by $T_m$ and $\eta(T_m)$. $f_1=0.55$ and $f_2=2.0$ are calibration coefficients that scale the thermodynamic and topological contributions to the final cooling temperature of the samples, while $Z\approx2\times10^{-6}$ is the empirical prefactor ~\cite{Takeuchi2001}. The melting temperature of the alloy is evaluated using the rule of mixtures,

\begin{equation}
	T_m=\sum_{i=1}^{n}c_iT_{m,i},
\end{equation}

where $T_{m,i}$ is the melting temperature of the $i$-th component. The effective viscosity $\eta(T_m)$ is calculated using the topology-dependent multiplier corresponding to the candidate lattice.

The numerical calculations were performed using tabulated elemental properties, viscosity data, and binary interaction parameters from the literature. Melting temperatures, molar volumes, atomic radii, and related elemental constants were taken from standard reference and review data~\cite{Gale2004,Miracle2017}. Elemental viscosities and their temperature dependences were evaluated using the tabulated data and correlations reported by Battezzati and Greer~\cite{Battezzati1989} and in the Smithells reference data~\cite{Gale2004}. Atomic-size and mixing-enthalpy parameters used in the glass-forming ability correction were based on the tabulations and empirical criteria of Takeuchi and Inoue~\cite{Takeuchi2005}, together with general phase-stability data summarized for multicomponent concentrated alloys by Miracle and Senkov~\cite{Miracle2017}.

For a given alloy composition, the critical cooling rate is calculated separately for the competing BCC-, FCC-, and, where relevant, HCP-like crystallization pathways. The phase with the highest calculated value of $R_c$ is identified as the kinetically dominant crystallization pathway. This interpretation follows from the physical meaning of $R_c$: a higher cooling rate is required to suppress a phase that nucleates and grows more readily under the same thermal conditions.

The criterion differs from equilibrium phase-stability rules. Static thermodynamic or electronic parameters identify phases favored under equilibrium or near-equilibrium conditions, whereas the present kinetic criterion identifies the structure that is most difficult to suppress during rapid heat extraction. Therefore, a metastable phase may become dominant if it has the lowest effective kinetic barrier at the nucleation stage.

Close $R_c$ values for different candidate structures indicate strong kinetic competition. In such cases, small variations in local composition, thermal gradient, cooling rate, or nucleation-site distribution may change the crystallization sequence, leading to multiphase or structurally heterogeneous solidification. Conversely, if the imposed cooling rate exceeds the critical cooling rates of all crystalline pathways, crystallization may be suppressed and an amorphous state may form, provided that the alloy also satisfies the glass-forming requirements.

The model is intended primarily for evaluating kinetic competition between BCC-, FCC-, and HCP-like crystallization pathways leading to simple solid-solution structures. In its present form, the model does not explicitly account for the formation of intermetallic compounds with complex crystal structures that are not derivatives of BCC or FCC lattices, such as topologically close-packed Laves phases or $\sigma$ phases. Therefore, the predicted dominance of a given crystalline pathway should be interpreted as the kinetic preference for the corresponding simple lattice during the early stage of crystallization, rather than as a complete description of the final phase constitution.

\section{Results and Discussion}

\subsection{Comparison with Experimental and Computational Data}

To test the model, its predictions were compared with experimental and computational data obtained in our previous studies of rapidly solidified and thin-film multicomponent alloys~\cite{KushnerovStaticCriteria,Kushnerov2021CoCrCuFeNi,Kushnerov2023MetastableFilms,Kushnerov2021ThinFilm,Kushnerov2025CantorThinFilm}. The selected examples illustrate changes in the lattice type expected from the valence electron concentration (VEC) criterion, kinetic suppression of phase separation, and competitive multiphase crystallization.

For each selected composition, the calculated critical cooling rates of the candidate BCC-, FCC-, and, where applicable, HCP-like pathways were compared using the kinetic selection criterion formulated above. The phase with the highest value of $R_c$ was treated as the kinetically dominant pathway, whereas close $R_c$ values were interpreted as an indication of competitive crystallization.

The first example concerns the competition between static electronic criteria and kinetic phase ranking. The VEC criterion is one of the most widely used parameters for estimating lattice preference in multicomponent alloys. In many systems based on transition metals, high VEC values are associated with FCC stabilization, whereas lower VEC values favor BCC-like structures. However, this criterion is essentially equilibrium-oriented and does not explicitly account for kinetic limitations during rapid solidification.

This limitation is illustrated by comparing the baseline CuFeMnNi alloy with the Al- and Si-containing alloy Al$_{0.5}$CuFeNiSi$_{0.25}$. For CuFeMnNi, the calculated VEC is 9.0, which is consistent with the conventional expectation of FCC stabilization. The kinetic model gives $R_c^{FCC}=5.786\times10^8$~K/s and $R_c^{BCC}=3.57\times10^8$~K/s, indicating FCC kinetic dominance. This agrees with the experimentally observed FCC single-phase state in rapidly quenched samples.

A different behavior is observed for Al$_{0.5}$CuFeNiSi$_{0.25}$. Its VEC value is 8.4, which still lies in the nominal FCC-favored range. However, the calculated critical cooling rate for the BCC pathway, $R_c^{BCC}=3.89\times10^7$~K/s, exceeds the FCC value, $R_c^{FCC}=3.344\times10^7$~K/s. Thus, the kinetic model predicts a change in the expected lattice preference. This result can be attributed to the combined topological influence of Al and Si, which modifies the atomic size distribution and changes the effective viscosity associated with different local structural arrangements.

The experimental phase constitution of rapidly quenched Al$_{0.5}$CuFeNiSi$_{0.25}$ contains both FCC and BCC-related phases. The presence of the BCC fraction is therefore consistent with the predicted BCC kinetic preference. This comparison shows that, in alloys containing strong topological modifiers, kinetic accessibility at the nucleation stage may override the lattice preference inferred from VEC alone.

A second group of examples concerns alloys in which positive mixing enthalpies promote phase separation. In multicomponent alloys containing element pairs such as Cu--Fe and Cu--Cr, equilibrium or near-equilibrium solidification commonly promotes chemical partitioning and secondary-phase formation. Under rapid solidification, the extent to which this tendency can be suppressed depends on the balance between nucleation kinetics and the available time for diffusion.

This behavior is illustrated by the Fe-rich alloy Fe$_5$CrCuNiMnSi and the Cu-rich alloy Cu$_5$CrFeMnNiSi. For Fe$_5$CrCuNiMnSi, macroscopic liquation is observed under slow cooling. The kinetic model gives $R_c^{FCC}=2.124\times10^7$~K/s and $R_c^{BCC}=1.387\times10^7$~K/s, indicating FCC kinetic dominance. After splat quenching, the alloy forms a supersaturated FCC solid solution with a lattice parameter $a=0.3615$~nm. This behavior is consistent with kinetic suppression of phase separation: rapid cooling suppresses macroscopic chemical partitioning and limits the growth of equilibrium secondary phases.

For Cu$_5$CrFeMnNiSi, the calculated values are higher and close to each other: $R_c^{FCC}=1.271\times10^8$~K/s and $R_c^{BCC}=8.406\times10^7$~K/s. The FCC pathway remains kinetically favored, but the BCC-related pathway is sufficiently competitive. This near competition corresponds to the experimentally observed multiphase structure of the rapidly quenched alloy, which contains an FCC matrix with $a=0.3649$~nm together with BCC-related precipitates represented by two B2-type phases with lattice parameters $a=0.2889$~nm and $a=0.2823$~nm.

The comparison between these two systems demonstrates the practical limit of kinetic suppression of phase separation. When one crystallization pathway is clearly dominant, rapid solidification can stabilize a supersaturated solid solution even in alloys with a thermodynamic tendency toward phase separation. When competing pathways have comparable critical cooling rates, residual chemical driving forces and local compositional fluctuations can still promote secondary-phase formation. In this regime, the model should be interpreted as identifying the likelihood of competitive solidification rather than predicting a strictly single-phase state.

\subsection{Glass Formation and Model Limitations}

The next group concerns alloys located near the boundary between glass formation and crystallization. These systems are useful for evaluating the role of the glass-forming ability correction because they contain elements commonly associated with amorphization, while their actual behavior depends on the balance between thermodynamic frustration, atomic size mismatch, and local chemical affinity.

The Nb-containing alloy FeCo$_{0.854}$Nb$_{0.146}$NiB$_{0.7}$Si$_{0.3}$ represents a composition with strong topological and thermodynamic conditions favorable for amorphization. The large atomic size mismatch introduced by Nb, combined with the presence of B and Si, increases packing frustration and enhances the glass-forming tendency. In the model, these effects lead to a high value of $GFA_{index}$ and, consequently, to a correction factor $F_{GFA}$ close to unity. The calculated critical cooling rate is low, $R_c\approx3.854\times10^2$~K/s, which is consistent with the experimentally observed formation of an amorphous phase after rapid solidification.

A contrasting behavior is observed when Nb is replaced by Be in FeCoNiB$_{0.7}$Si$_{0.3}$Be. Although this alloy also contains B and Si, the substitution changes the balance of thermodynamic and topological factors. The glass-forming ability index decreases, and the correction factor increases accordingly. As a result, the critical cooling rate for the dominant crystalline pathway rises to $R_c\approx1.04\times10^4$~K/s. Experimentally, the rapidly quenched alloy crystallizes into a BCC-related ordered B2 phase accompanied by secondary borides. Thus, the model captures the trend that the Be-containing alloy is a weaker glass former than the Nb-containing composition.

This example also illustrates an important limitation of the macroscopic approach. The model describes crystallization through averaged thermodynamic, topological, and viscosity-related quantities. It does not explicitly account for chemically specific short-range-order clusters in the liquid. In systems with strong local chemical affinity, such clusters may act as effective precursors for ordered phases or borides, reducing the effective nucleation barrier beyond what is captured by the macroscopic viscosity-based description.

An additional test of the kinetic ranking is provided by atomistic molecular dynamics simulations of the earliest stages of structural ordering during thin-film deposition. Such simulations provide direct information about transient nuclei and local ordering processes under strongly non-equilibrium conditions, where experimental identification of the first crystallization events is extremely difficult. To examine the robustness of the kinetic approach, two different thin-film deposition cases were considered.

The first case concerns the equiatomic AlCoCuFeNi alloy, for which the prediction based on the valence electron concentration criterion differs from the non-equilibrium structural ordering observed during deposition. According to the conventional VEC criterion, this alloy has $\mathrm{VEC}\approx8.2$, which suggests FCC stabilization. However, the kinetic model predicts a different ordering sequence. The calculated critical cooling rate for the BCC-like pathway is $R_c^{BCC}=4.69\times10^7$~K/s, whereas for the FCC-like pathway it is lower, $R_c^{FCC}=3.96\times10^7$~K/s. Thus, the macroscopic model predicts BCC kinetic priority at the earliest stage of ordering.

This prediction is consistent with atomistic molecular dynamics simulations of ultra-rapid non-equilibrium growth during the deposition of AlCoCuFeNi thin films~\cite{Kushnerov2021ThinFilm}. The simulations show that the initial stages of film growth are characterized by the preferential formation of BCC-like clusters at the substrate interface. As deposition proceeds and the film thickness increases, accumulated local stresses and chemical-size distortions promote partial structural rearrangement. As a result, the growing film evolves toward a heterogeneous multiphase state containing highly faulted FCC- and HCP-like regions together with retained BCC-like domains.

The second case examines dynamic compositional effects during deposition of the equiatomic CoCrFeMnNi alloy onto an Al substrate~\cite{Kushnerov2025CantorThinFilm}. For the nominal CoCrFeMnNi composition, both conventional static criteria and the present kinetic model predict FCC dominance. However, molecular dynamics simulations indicate that the first nanolayers of the growing film can crystallize into a BCC-like structure. This apparent contradiction is resolved when local compositional changes near the substrate are taken into account. During the initial deposition stage, Al atoms from the substrate diffuse into the growing film, locally increasing the Al concentration. Recalculation of the kinetic parameters for this Al-enriched local composition leads to a pronounced increase in $R_c^{BCC}$ and gives the BCC-like pathway a transient kinetic priority. As the film becomes thicker and the Al concentration decreases toward the nominal CoCrFeMnNi composition, the kinetic preference shifts back toward the FCC-like pathway.

The agreement between the macroscopic kinetic ranking and the atomistically observed ordering sequences in both fixed-composition and dynamically changing local-composition cases supports the central assumption of the present framework. Under strongly non-equilibrium conditions, the first structural state to appear is governed by kinetic priority at the nucleation stage rather than by equilibrium lattice preference alone. The atomistic results also show that the proposed macroscopic model is sensitive to the local chemical environment and can be used to identify the dominant early crystallization pathway, provided that the relevant local composition is used in the calculation.

Taken together, the selected examples show that the model is most reliable when used to evaluate kinetic competition between BCC-, FCC-, and HCP-like crystallization pathways leading to simple solid-solution structures during the early stages of solidification. Its predictions should therefore be interpreted primarily in terms of BCC-, FCC-, or HCP-like crystallization pathways. In its present form, the model does not explicitly describe chemical partitioning, long-range diffusion, chemically specific short-range order, or the formation of intermetallic compounds with complex crystal structures that are not derivatives of BCC or FCC lattices, such as topologically close-packed Laves phases or $\sigma$ phases.

These limitations define the natural scope of the present semi-empirical framework. The model enables preliminary evaluation of kinetic phase-selection tendencies under rapid cooling and is not intended to replace detailed atomistic simulations, CALPHAD calculations, phase-field modeling, or experimental structural analysis. In systems where local chemical affinity strongly affects phase selection, further development of the approach should include parameters describing chemical short-range order, phase-specific partitioning, and nucleation involving locally chemically or structurally ordered regions.

\section{Conclusions}

A semi-empirical kinetic framework for predicting phase selection in rapidly solidified multicomponent concentrated alloys has been formulated. In contrast to static thermodynamic and electronic criteria, the proposed approach treats phase formation as a kinetic competition between alternative crystallization pathways. The critical cooling rate is used as a quantitative criterion for ranking the kinetic priority of competing BCC-, FCC-, and HCP-like crystallization pathways.

A distinctive feature of the framework is the incorporation of a topology-dependent viscosity correction into the critical-cooling-rate criterion. This correction accounts for differences in local atomic packing associated with different simple lattice types. To avoid the non-physical discontinuity produced by the conventional piecewise BCC topological multiplier near the critical atomic-size-mismatch threshold, a smoothed BCC multiplier based on hyperbolic-tangent weighting functions was introduced. The proposed regularization stabilizes the calculated viscosity and reduces artificial jumps in the predicted critical cooling rates.

An additional original aspect of the model is the glass-forming ability correction factor $F_{GFA}$ based on the index $GFA_{index}$. This correction reflects the fact that efficient amorphization requires simultaneous thermodynamic and topological frustration. The use of smooth threshold score functions provides a continuous transition between alloys with low and high glass-forming ability and prevents underestimation of the critical cooling rate for weak glass-forming alloys.

Comparison with previously published experimental and computational data shows that the model reproduces the main phase-selection trends observed under rapid quenching and non-equilibrium thin-film deposition. These trends include changes in the lattice type expected from the valence electron concentration criterion, kinetic suppression of phase separation in systems with positive mixing enthalpies, competitive multiphase crystallization, and the different behavior of alloys with high and low glass-forming ability.

The proposed model enables preliminary evaluation of which BCC-, FCC-, or HCP-like crystalline states are kinetically favored during the early stages of solidification. It should therefore be interpreted as an approach for identifying early crystallization tendencies rather than as a complete description of the final phase constitution. When chemical partitioning, short-range order, or complex intermetallic phases play an important role, the model results should be complemented by thermodynamic, atomistic, or experimental analysis.

\end{document}